\documentclass[a4paper]{jpconf}
\usepackage{graphicx}
\usepackage{amsmath,amssymb,bm}

\begin{document}
\title{Higher order fluctuations of conserved charges in heavy ion collisions
}

\author{Miki Sakaida, Masayuki Asakawa, Masakiyo Kitazawa}

\address{Department of Physics, Osaka University, Toyonaka, Osaka 560-0043, Japan}

\ead{sakaida@kern.phys.sci.osaka-u.ac.jp}

\setlength{\abovedisplayskip}{6pt} % 上部のマージン
\setlength{\belowdisplayskip}{6pt} % 下部のマージン

\begin{abstract}

We investigate the effect of the global charge conservation 
on the cumulants of conserved charges in relativistic 
heavy ion collisions in a finite rapidity window 
by studying the time evolution of cumulants 
in the hadronic medium.
It is argued that the global charge conservation does not affect the experimental result of the net-electric 
charge fluctuation observed by ALICE, because of the finite 
diffusion distance of charged particles in
the hadronic stage.

\end{abstract}

\section{Introduction}

Bulk fluctuations of conserved charges measured by event-by-event analyse
s in relativistic heavy ion collisions, especially higher order cumulants, are expected to be promising observables to characterize 
properties of the hot medium created by collision events 
and to find the QCD critical point
\cite{Koch:2008ia,Asakawa:2000wh,Kitazawa:2013bta}. 
Recently, experimental investigations of these observables 
in heavy ion collisions have been actively performed at the 
RHIC \cite{STAR} and LHC \cite{ALICE}. 

One of the important properties of these fluctuations in heavy ion collisions is that experimentally-observed fluctuations had experienced non-equilibrium evolution, because of the dynamical nature of the hot medium created in heavy ion collisions  \cite{Kitazawa:2013bta,Shuryak:2000pd}.
In fact, the recent experimental result on the rapidity window dependence of net electric charge fluctuation 
by the ALICE Collaboration at the LHC \cite{ALICE} shows that the value 
of the second order cumulant of the net electric charge is significantly suppressed 
compared with the one in the equilibrated hadronic medium. 
This result is reasonably understood if one interprets the 
suppression as a sign of the survival of the small fluctuation created in 
the primordial deconfined medium \cite{Asakawa:2000wh,Kitazawa:2013bta,
Shuryak:2000pd}.

However, there exists another mechanism 
called the global charge conservation (GCC)
to cause the suppression
of fluctuations compared with the 
equilibrated value. 
If one counts a conserved charge in the total system
generated by experiments,
there are no event-by-event fluctuations because 
of the charge conservation.
This fact is referred to as the GCC
\cite{Koch:2008ia}.
In Ref.~\cite{Bleicher:2000ek}, 
on the assumption that the equilibration is established in 
the final state of the heavy ion collision,
the magnitude of this effect 
is estimated.
It, however, should be noted that 
the suppression of charge fluctuation observed at ALICE \cite{ALICE}
is stronger than the one estimated in Ref.~\cite{Bleicher:2000ek}.
This result shows that there exists another contribution
for the suppression in addition to the GCC, such as 
the nonequilibrium effect
as originally addressed in Ref.~\cite{Asakawa:2000wh}.
The effect of the GCC at LHC energy, therefore, has to be 
discussed again with the nonthermal effects incorporated.

In the present study, we investigate the effect of the 
GCC on cumulants of conserved charges 
under such nonequilibrium circumstances, by describing the time
evolution of fluctuations in a finite volume system \cite{Sakaida:2014pya}.
We also discuss the effects of the GCC
on higher-order cumulants of conserved charges for the 
first time.
By comparing our result with the experimental
one at ALICE in Ref.~\cite{ALICE}, we find that the effect of GCC on
the net electric charge fluctuation in the rapidity window observed 
by this experiment is almost negligible.

\section{Stochastic formalism to describe diffusion of hadrons}
In heavy ion collisions with sufficiently large $\sqrt{s_{\rm NN}}$,
the hot medium created at the mid-rapidity has an approximate boost 
invariance.
We denote the net number of a conserved charge per 
unit space-time rapidity $\eta$ as $n(\eta,\tau)$ with the proper time $\tau$, and 
set $\tau=\tau_0$ at hadronization. 
Because of the local charge conservation, 
the probability distribution of $n(\eta,\tau_0)$ at hadronization 
inherits from the one that existed in 
the deconfined medium \cite{Koch:2008ia}. 
After hadronization, particles diffuse 
and rescatter, and the distribution of $n(\eta,\tau)$ 
continues to approach the one of the equilibrated hadronic medium
until kinetic freezeout at $\tau = \tau_{\rm fo}$.

In this study, to describe the time evolution of fluctuation of the particle number at mid-rapidity $Q(\Delta\eta,~\tau)= \int_{-\Delta\eta/2}^{\Delta\eta/2} d\eta \ n(\eta,\tau)$ in hadronic medium, with the size of the rapidity window to count the charge number $\Delta\eta$, until $\tau = \tau_{\rm fo}$ we adopt the diffusion master equation \cite{Kitazawa:2013bta,Sakaida:2014pya}.
In this model, we divide the system with the total rapidity length 
$\eta_{\rm tot}$ into $M$ discrete cells with an equal finite 
length $a=\eta_{\rm tot}/M$. We then consider a single species 
of particles for the moment, and denote the particle number 
existing in the $m$th cell as $n_m$ and the probability distribution 
that each cell contains $n_m$ particles as $P(\bm{n}, \tau)$ with 
$\bm{n}=(n_{0},n_{1}, \cdots ,n_{m}, \cdots ,n_{M-2},n_{M-1} )$.
Finally, we assume that each particle moves to the adjacent cells 
with a probability $\gamma (\tau)$ per unit proper time, 
as a result of microscopic interactions. 
The probability $P(\bm{n},\tau)$ then obeys the diffusion master equation,
\begin{align}
\partial_\tau P(\bm{n},\tau)
= \gamma(\tau) \sum_{m=0}^{M-1}[( n_{m} + 1 ) 
\{ P(\bm{n}+\bm{e}_{m}-\bm{e}_{m+1},\tau)+ P(\bm{n} +\bm{e}_{m}-\bm{e}_{m-1},\tau) \} 
- 2 n_m P(\bm{n},\tau) ]\bigr],
\label{eq:DME}
\end{align}
where $\bm{e}_{m}$ is a unit vector whose all components are 
zero except for $m$th one, which takes unity.

The average and the Gaussian fluctuation of $n(\eta,\tau)$ 
in Eq.~(\ref{eq:DME}) in the continuum limit, $a\to0$, agree 
with the ones of the stochastic diffusion equation, 
\cite{Shuryak:2000pd}
\begin{eqnarray}
\partial_\tau n(\eta,\tau) 
= D(\tau) \partial_\eta^2 n(\eta,\tau)+\partial_\eta \xi(\eta,\tau) ,
\label{eq:diffusion}
\end{eqnarray}
when one sets the diffusion coefficient as $D(\tau)=\gamma(\tau)a^2$ \cite{Kitazawa:2013bta}. Here, $\xi(\eta,\tau)$ is the temporarily-local 
stochastic force.
It is known 
that the fluctuation of
Eq.~(\ref{eq:diffusion}) in equilibrium becomes of Gaussian 
\cite{Kitazawa:2013bta,Sakaida:2014pya}.
On the other hand, the diffusion master equation
Eq.~(\ref{eq:DME}) can give rise to nonzero higher order cumulants 
in equilibrium \cite{Kitazawa:2013bta}.
This is the reason why we employ the diffusion master equation
instead of Eq.~(\ref{eq:diffusion}).

\section{Solution of diffusion master equation}
Now, we determine the time evolution of cumulants 
by solving Eq.~(\ref{eq:DME}) and taking the continuum limit $a\to 0$. 
In order to take account of the finiteness of the size of the hot medium, we 
set reflecting boundaries at $\eta=\pm\eta_{\rm tot}/2$.
After some algebra \cite{Sakaida:2014pya}, 
the cumulants of $Q(\tau)$ with the fixed initial condition $n(\eta,\tau_0)=N(\eta)$ 
are calculated to be
\begin{eqnarray}
\langle (Q(\Delta\eta,~\tau))^n \rangle_{\rm c} 
= \int_{-\eta_{\rm tot}/2}^{\eta_{\rm tot}/2} d\eta \ N(\eta) H_n(\eta) ,
\label{eq:<Q^n>}
\end{eqnarray}
where 
 $H_2(\eta) = I(\eta) - I(\eta)^2 , H_4(\eta) = 
I(\eta) - 7 I(\eta)^2 + 12 I(\eta)^3- 6 I(\eta)^4 $ with
\begin{eqnarray}
I(\eta)=\frac{\Delta\eta}{\eta_{\rm tot}}\sum_{k=-\infty}^{\infty} \cos \left( \frac{\pi k \eta}{\eta_{\rm tot}}\right) \sin \left({\displaystyle \frac{\pi k\Delta\eta}{2\eta_{\rm tot}} }\right) \cos \left(\frac{\pi k}{2}\right) \exp\left[-\frac{1}{2}\left(\frac{\pi k d(\tau)}{\eta_{\rm tot}}\right)^2\right]/ \left( {\displaystyle \frac{\pi k\Delta\eta}{2\eta_{\rm tot}} }\right) .
\label{eq:I_X}
\end{eqnarray}
Here, 
$d(\tau)=\sqrt{2\int_{\tau_0}^{\tau}d\tau'D(\tau')}$ 
is the average diffusion length of each Brownian particle.

Next, we extend the above results to the 
cases with general initial conditions with non-vanishing initial fluctuations. 
In the following, we also extend the formula to treat
cumulants of the difference of densities of two particle species, $n_1(\eta,\tau)$ and $n_2(\eta,\tau)$, $Q_{\rm (net)}(\Delta\eta,~\tau)=\int_{-\Delta\eta/2}^{\Delta\eta/2} d\eta \ (n_1(\eta,\tau)-n_2(\eta,\tau))$,
in order to consider cumulants of conserved charges in QCD.

Next, let us constrain the initial condition.
Because we consider a finite system, 
the initial condition at $\tau=\tau_0$
should be determined in accordance with the GCC.
In order to obtain the initial conditions 
which conform the GCC,
here we model the initial configuration by
an equilibrated free classical gas in a finite volume.
In this system,
one can obtain the correlation functions by taking the 
continuum limit of the multinomial distribution \cite{Sakaida:2014pya}.
On the other hand, we assume that 
the total charge number at $\tau=\tau_0$ in the total system 
obeys the one given by the grand canonical ensemble \cite{Kitazawa:2013bta}.
Using these initial conditions and Eq.~(\ref{eq:<Q^n>}),
one obtains the higher order cumulants of 
$Q_{\rm(net)}(\Delta\eta,~\tau)$.

\section{Effects of the GCC}

Next, let us study how the GCC affects
the rapidity window dependence of the cumulants of conserved 
charges.
To describe the time evolution of cumulants, we use 
the dimensionless parameters $T \equiv d(\tau)/\eta_{\rm tot}$.
To make the argument simple, we consider the $\Delta\eta$ dependence for 
the fixed initial condition,
i.e., all fluctuations vanish at $\tau=\tau_0$.
In Fig.~1, 
we show $\langle (Q_{{\rm (net)}}  (\Delta\eta))^n \rangle_{\rm c}
/\langle Q_{{\rm (tot)}}  (\Delta\eta) \rangle_{\rm c}$ for $n=2$ (left) and $4$ (right), as a function of $\Delta\eta/\eta_{\rm tot}$
with the total length $\eta_{\rm tot}$ for five values of $T$ including infinity. 
For comparison, we also show the result with an infinite volume 
by the dashed and dotted lines.
The figure shows that $\langle (Q_{{\rm (net)}}  (\Delta\eta))^2 \rangle_{\rm c}$ 
vanishes at $\Delta\eta/\eta_{\rm tot}=1$ for each $T$, namely 
$\Delta\eta=\eta_{\rm tot}$, which is a trivial consequence of 
the GCC.
On the other hand, as $\Delta\eta$ becomes smaller from this value with fixed $T$,
the result  approaches the one with infinite 
volume. The effect of the GCC vanishes almost completely 
except for the range, 
\begin{align}
\frac{\eta_{\rm tot}- \Delta \eta}{2} \lesssim d(\tau).
\label{eq:effec}
\end{align}
In this expression, the left-hand side is the distance between 
the left (right) boundary and the left (right) edge of the rapidity window, while the 
right-hand side is the diffusion length of each Brownian particle.
When Eq.~(\ref{eq:effec}) is satisfied, particles which are reflected
by one of the boundaries at least once can enter the rapidity window.
Therefore, the existence of the boundaries can 
affect the fluctuations of conserved charges in the rapidity window.
On the other hand, when the condition Eq.~(\ref{eq:effec}) 
is not satisfied, particles inside the rapidity window do not 
know the existence of the boundaries; in other words, 
the fact that the system is finite.
In the latter
case, therefore, the fluctuations in the rapidity window
are free from the effect of the GCC.

With fixed $\eta_{\rm tot}$, larger $T$ corresponds to larger $\tau$.
Figure~1(left) shows that as $T$ becomes larger 
the fluctuation increases and approaches a linear function 
representing the equilibration, which is consistent with the result obtained in Ref.~\cite{Bleicher:2000ek}.

From Fig.~1 (right), which shows the 
fourth order cumulants for several values of $T$,
one obtains completely the same conclusion on the effect of
the GCC: The effect of the GCC
is visible only for large values of $\Delta\eta/\eta_{\rm tot}$.

In this analysis, it is assumed that the particle current vanishes
on the two boundaries. This assumption would not be suitable to 
describe the hot medium created by heavy ion collisions, 
since the hot medium does not have such hard boundaries but
only baryon rich regions.
From the above discussion, however, it is obvious that
the effect of boundaries does not affect the fluctuations in 
the rapidity window unless the condition Eq.~(\ref{eq:effec}) 
is satisfied irrespective of the types of the boundaries.

\begin{figure}
\begin{minipage}{0.5\hsize}
        \begin{center}       
          \includegraphics[keepaspectratio, angle=-90, clip, width=8.5cm]{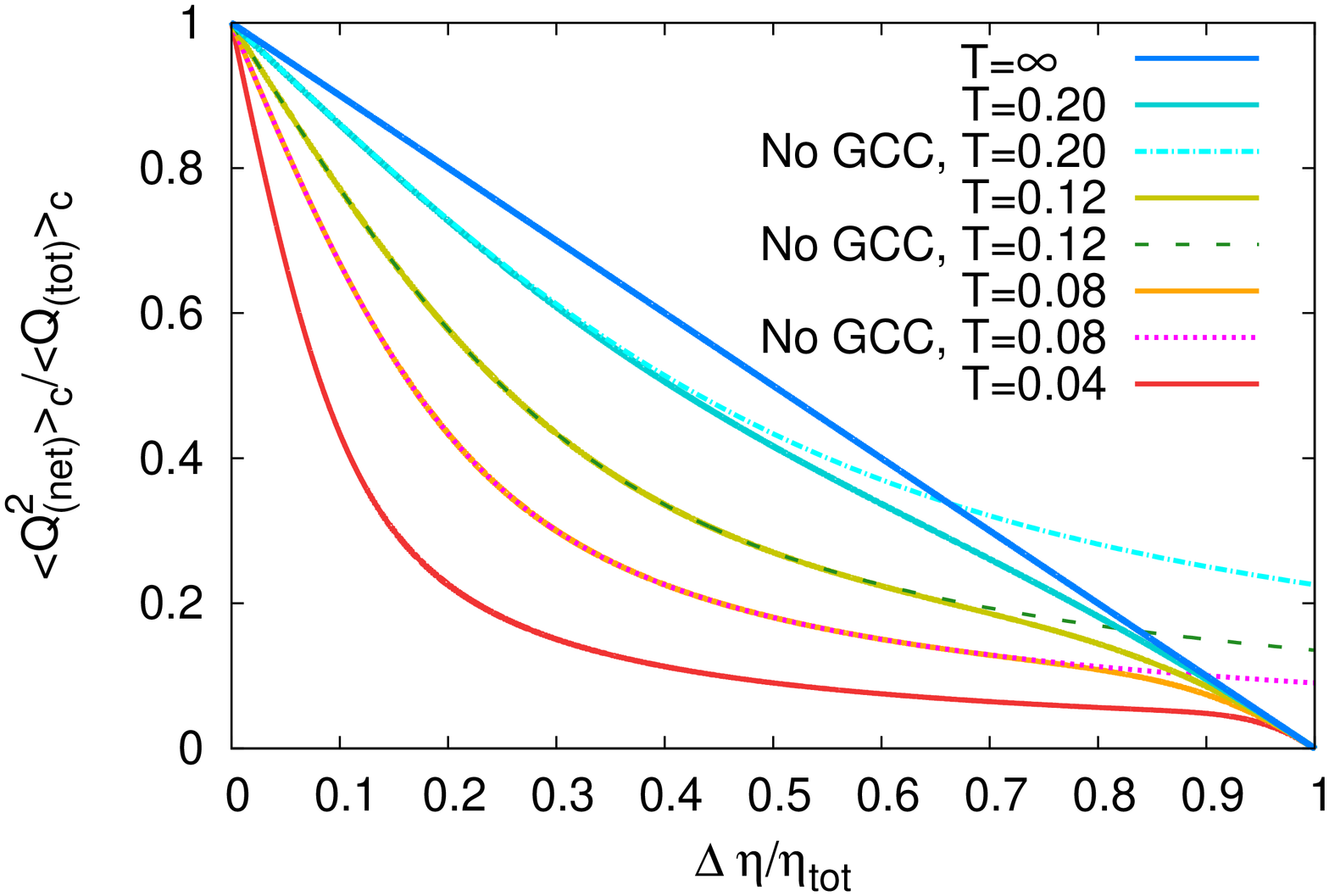}
             \label{fig:22b0}
      \end{center}
 \end{minipage}
 \begin{minipage}{0.5\hsize}
      \begin{center}       
          \includegraphics[keepaspectratio, angle=-90, clip, width=8.7cm]{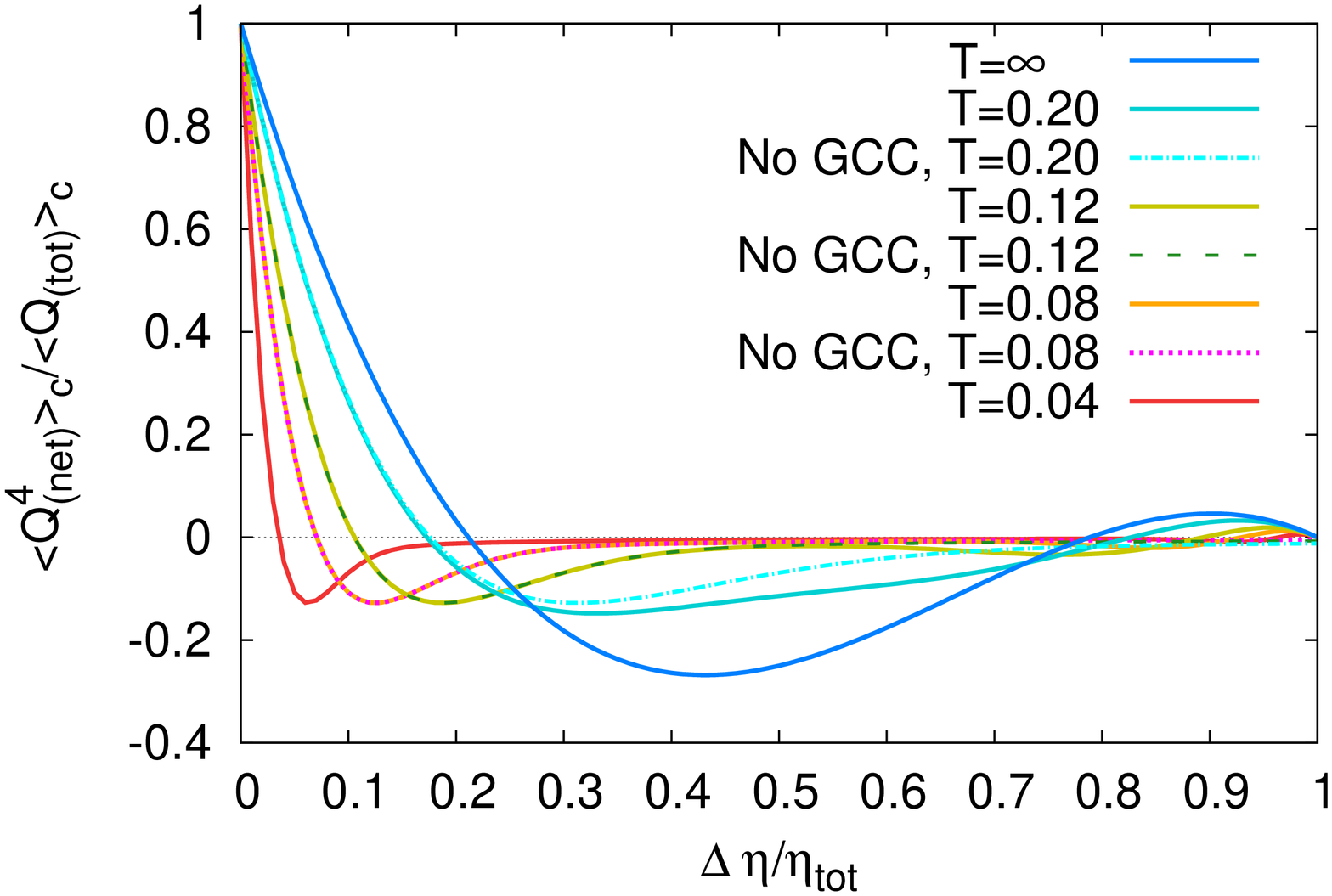}
             \label{fig:4th20}  
      \end{center}
       \end{minipage}
       \vspace*{-0.6cm} 
       \caption{
Second (left panel) and forth (right panel) order cumulants of conserved charges without initial 
fluctuation as a function of $\Delta\eta/\eta_{\rm tot}$ 
with five values of the parameter $T$.
The corresponding results in an infinite volume are also plotted
for several values of $T$.
}
\end{figure}

We finally inspect the $\Delta\eta$ dependence
of the net electric charge fluctuations observed at ALICE \cite{ALICE}
in more detail.
In order to estimate the effect of the GCC, 
we must determine the magnitude of $\eta_{\rm tot}$.
From the pseudo-rapidity dependence of charged-particle yield 
at LHC energy, $\sqrt{s_{\rm NN}}=2.76$ TeV, in Ref.~\cite{Abbas:2013bpa}, 
we take the value $\eta_{\rm tot}=8$ in the following.
With this $\eta_{\rm tot}$, and  
the maximum rapidity coverage of the $2\pi$ detector, TPC, of ALICE, 
$\Delta\eta=1.6$ \cite{ALICE}, one obtains 
$\Delta\eta/\eta_{\rm tot} = 0.2$.
From Fig.~1, 
one can immediately conclude that 
the suppression of $\langle (Q_{\rm (net)} (\Delta\eta))^2 \rangle_c$ in 
this experiment cannot be explained solely by the naive formula 
of the GCC \cite{Bleicher:2000ek}.
From the discussion in the previous subsections,
it is also concluded that the effect of the GCC
on the diffusion in the hadronic stage is negligible in this 
experimental result.

\end{document}